\documentclass[useAMS,usenatbib,onecolumn]{mn2e}
\usepackage{graphics,graphicx,epsf}

\title[Negative skewness of radial pairwise velocity]
{Negative skewness of radial pairwise velocity in the quasi-nonlinear regime: Zel'dovich approximation}

\author[Yoshisato et al.]
{Ayako Yoshisato$^{1,2}$\thanks{E-mail:
ayako@cosmos.phys.ocha.ac.jp, yosisisato@isp.co.jp (AY); hiro@phys.ocha.ac.jp (MM); hmouri@mri-jma.go.jp (HM)}, 
Masahiro Morikawa$^{1}$\footnotemark[1] and 
Hideaki Mouri$^{3}$\footnotemark[1]\\
$^{1}$Department of Physics, Ochanomizu University, Otsuka 2-2-1, Bunkyo, Tokyo 112-0012, Japan\\
$^{2}$Research Institute of Systems Planning, Sakuragaoka-cho 2-9, Sibuya, Tokyo 150-0031, Japan\\
$^{3}$Meteorological Research Institute, Nagamine 1-1, Tsukuba 305-0052, Japan}
\begin{document}

\date{Accepted 2003 May 1; in original form 2002 October 10}

\pagerange{\pageref{firstpage}--\pageref{lastpage}} \pubyear{2003}

\maketitle

\label{firstpage}

\begin{abstract}
According to $N$-body numerical simulations, the radial pairwise velocities of galaxies have negative skewness in the quasi-nonlinear regime. To understand its origin, we calculate the probability distribution function of the radial pairwise velocity using the Zel'dovich approximation, i.e., an analytical approximation for gravitational clustering. The calculated probability distribution function is in good agreement with the result of $N$-body simulations. Thus the negative skewness originates in relative motions of galaxies in the clustering process that the infall dominates over the expansion. 
\end{abstract}

\begin{keywords}
cosmology: theory --- large-scale structure of universe
\end{keywords}

\section{INTRODUCTION}

Let us consider two galaxies separated from each other by the vector ${\bmath r}$. If their peculiar velocities are ${\bmath u}_1$ and ${\bmath u}_2$, the pairwise velocity ${\bmath v}$ is defined as 
\begin{eqnarray}
{\bmath v} = {\bmath u}_2 - {\bmath u}_1 
= v_{\parallel} {\bmath n}_{\parallel} + v_{\perp} {\bmath n}_{\perp}.
\label{intro1}
\end{eqnarray}
Here the unit vectors ${\bmath n}_{\parallel}$ and ${\bmath n}_{\perp}$ are parallel and perpendicular to the separation vector ${\bmath r}$. We study probability distribution functions (PDFs) of $v_{\parallel}$ and $v_{\perp}$, which are respectively defined as radial and transverse components of the pairwise velocity ($-\infty < v_{\parallel} < +\infty$, $0 \le v_{\perp} < +\infty$).

The radial pairwise velocity is of fundamental importance in observational cosmology (Peebles 1993, section 20). It serves as a probe of the matter distribution. Its PDF is essential to converting observed data in the redshift space into those in the real space.

For the quasi-nonlinear regime of the present-day universe ($r = |{\bmath r}| \simeq 10^1$--$10^2$ Mpc), $N$-body numerical simulations have shown that the radial-velocity PDF exhibits a negative average and negatively skewed pronounced tails (Efstathiou et al. 1988; Fisher et al. 1994; Zurek et al. 1994; Magira, Jing \& Suto 2000). These properties are attributable to coherent motions of galaxies approaching each other, since gravitational clustering is the predominant process in the quasi-nonlinear regime.

However, with the results of {\it N}-body simulations alone, we cannot understand how the negative average, negative skewness and pronounced tails of the radial-velocity PDF are related with gravitational clustering. To clarify this relation, an analytical treatment is desirable.

The linear perturbation, i.e., the most convenient analytic tool, in insufficient to study the above-mentioned properties of the radial-velocity PDF. This is because these properties are nonlinear. The linear perturbation merely amplifies peculiar velocities of all the galaxies by the same factor. Thus the radial-velocity PDF retains its Gaussianity in the initial stage, where motions of the individual galaxies are random and independent. Although the leading terms of the average $\langle v_{\parallel} \rangle$ and standard deviation $\langle v_{\parallel}^2 \rangle$ happen to be squares of the linear perturbation, those of the higher-order moments, e.g., $\langle v_{\parallel}^3 \rangle$ and $\langle v_{\parallel}^4 \rangle$, are nonlinear perturbations (see Juszkiewicz, Fisher \& Szapudi 1998).

Juszkiewicz et al. (1998) used a second-order Eulerian perturbation and reproduced successfully the moments up to the third order, $\langle v_{\parallel} \rangle$, $\langle v_{\parallel}^2 \rangle$, and $\langle v_{\parallel}^3 \rangle$, as well as the negatively skewed PDF. However, the PDF was obtained with a conventional but unjustified ansatz, instead of the rigorous theory that requires complicated calculations. Seto \& Yokoyama (1998) used a first-order Lagrangian perturbation, i.e., the Zel'dovich approximation (ZA; Zel'dovich 1970). Although their radial-velocity PDF exhibits pronounced tails, they failed to reproduce the negative average and had to shift their PDF by hand in order to achieve agreement with $N$-body simulations. Overall, the mechanism that determines the radial-velocity PDF is still controversial.

We use ZA to reexamine the PDF. It would be surprising if ZA really reproduced the pronounced tails but not the negative average $\langle v_{\parallel} \rangle$. The former is associated with the higher moments such as $\langle v_{\parallel}^4 \rangle$. For the quasi-nonlinear regime, ZA is a good approximation (Yoshisato, Matsubara \& Morikawa 1998). Each of the galaxies is assumed to move along the gravitational force that is determined by the initial density distribution. Thus the galaxies cluster together. The velocity and density fields undergo a nonlinear evolution, which is expected to cause the negative average, negative skewness and pronounced tails of the radial-velocity PDF.

We have to be careful about the extent to which ZA can describe gravitational clustering. Of particular importance is the range of the separation $r$ where ZA is applicable. It has been known that ZA is applicable only to a limited range, but the applicable range has not been estimated so far. The quantitative estimation is done for the first time in this paper. We subsequently demonstrate that Seto \& Yokoyama (1998) made their ZA analysis outside its applicable range.

Throughout the main text, we adopt the Einstein-de Sitter universe with a Hubble constant $H_{0} = 50$ km s$^{-1}$ Mpc$^{-1}$ for comparison with the previous works. In Section 2, an analytic form of the joint PDF for the radial and transverse velocities is derived with ZA. The applicable separation range of ZA is estimated in Section 3. We compare ZA with $N$-body simulations in Section 4. The origin of the negative average and negative skewness is discussed in Section 5. We discuss the existing relevant models in Section 6. The conclusion is presented in Section 7. The application to other cosmological models and so on are discussed in Appendices.

\section{PAIRWISE VELOCITY DISTRIBUTION}

\subsection{Analytical form}

Using ZA, we newly derive an analytic form of the joint PDF for the radial and transverse velocities at the separation $r$ and the time $t$, $P ( v_{\parallel} , v_{\perp} | r , t )$. It is assumed that the vectors $v_{\parallel} {\bmath n}_{\parallel}$ and $v_{\perp} {\bmath n}_{\perp}$ have isotropic distributions. The joint PDF is normalized as
\begin{eqnarray}
\label{norm}
\int P \left( v_{\parallel} , v_{\perp} | r , t \right)
d v_{\parallel} 2 \pi v_{\perp} d v_{\perp} 
= \xi(r,t) + 1,
\label{pdfnor2}
\end{eqnarray}
where $\xi (r, t)$ is the two-point correlation function for the number density of galaxies. The radial-velocity PDF, $P ( v_{\parallel}| r , t )$, is obtained by integrating the joint PDF over the transverse velocity $v_{\perp}$.

Let us consider a relative motion of two galaxies. Their separation vector ${\bmath r} = {\bmath r}(t)$ and pairwise velocity ${\bmath v} ={\bmath v}(t)$ are written with ZA as 
\begin{eqnarray}
\begin{array}{ll}
{\bmath r} = {\bmath x}_{2} - {\bmath x}_{1} = 
 {\bmath r}_{i} + \displaystyle \frac{D}{\dot{D_{i}}} {\bmath v}_{i} ,\\
{\bmath v} = {\bmath u}_{2} - {\bmath u}_{1} = 
             v_{\parallel} {\bmath n}_{\parallel} + v_{\perp} {\bmath n}_{\perp} =
 \displaystyle \frac{\dot{D}}{\dot{D_{i}}} {\bmath v}_{i}.
\label{pdf1}
\end{array}   
\end{eqnarray}
Here $D = D(t) \propto t^{2/3}$ is the linear growth factor of the density fluctuation. We set $D=1$ for the present-day universe. The suffix $i$ indicates quantities at the initial time $t = t_i$. Since the galaxies move along straight lines, the relative motion lies on the plane determined by the vectors ${\bmath r}_{i}$ and ${\bmath v}_{i}$.

With an increase of the time from $t$ to $t+dt$, a galaxy pair is defined to move from $(v_{\parallel}, v_{\perp}, r)$ to $(v_{\parallel}', v_{\perp}', r')$. From conservation of the number of galaxies, we have
\begin{equation}
P \left( v_{\parallel}' , v_{\perp}' | r' , t + dt \right) 
d v_{\parallel}' 2 \pi v_{\perp}' d v_{\perp}' 4 \pi r'^2 d r' 
= P \left( v_{\parallel} , v_{\perp} | r , t \right) 
d v_{\parallel} 2 \pi v_{\perp} d v_{\perp} 4 \pi r^2 d r.
\label{pdfeq1}
\end{equation}
From equation (\ref{pdf1}), the time evolution of $d v_{\parallel} 2 \pi v_{\perp} d v_{\perp} 4 \pi r^2 d r$ is obtained as
\begin{equation}
d v_{\parallel}' 2 \pi v_{\perp}' d v_{\perp}' 4 \pi r'^2 d r'
 = \left( 1 + 3 \displaystyle \frac{\ddot{D}}{\dot{D}} dt \right) 
d v_{\parallel} 2 \pi v_{\perp} d v_{\perp} 4 \pi r^2 d r.
\label{pdfeq10}
\end{equation}
The solution is 
\begin{eqnarray}
d v_{\parallel} 2 \pi v_{\perp} d v_{\perp} 4 \pi r^2 d r
 = \left( \frac{\dot{D}}{\dot{D}_{i}} \right)^3
d v_{\parallel,i} 2 \pi v_{\perp,i} d v_{\perp,i} 4 \pi r_{i}^2 d r_{i}.
\label{pdfeq11}
\end{eqnarray}
Thus the joint PDF at the time $t$ is related to the PDF at the initial time $t_i$ as
\begin{eqnarray}
P \left( v_{\parallel} , v_{\perp} | r , t \right)
 = \left( \frac{\dot{D}_{i}}{\dot{D}} \right)^3
 P \left( v_{\parallel,i} , v_{\perp,i} | r_{i} , t_{i} \right).
\label{pdfeq12}
\end{eqnarray}
From equation (\ref{pdf1}), the initial separation $r_i$ is obtained as 
\begin{eqnarray}
r_{i}=\sqrt{\left( r - \frac{D}{\dot{D}} v_{\parallel}
  \right)^2 + \left( \frac{D}{\dot{D}} v_{\perp} \right)^2}.
\label{pdf7} 
\end{eqnarray}
Likewise, the initial pairwise velocities $v_{\parallel,i}$ and $v_{\perp,i}$ are obtained as
\begin{eqnarray}
\begin{array}{ll}
v_{\parallel,i} = 
\displaystyle \frac{\dot{D_{i}}}{\dot{D}} 
\left( \displaystyle \frac{r v_{\parallel}}{r_{i}} - 
\displaystyle \frac{D}{\dot{D}} 
\displaystyle \frac{v_{\parallel}^2 + v_{\perp}^2}{r_{i}} \right) ,
\\
v_{\perp,i} = 
\displaystyle \frac{\dot{D_{i}}}{\dot{D}} \frac{r v_{\perp}}{r_{i}} ,
\label{pdf9}
\end{array}   
\end{eqnarray}
(see also Appendix A). Seto \& Yokoyama (1998) used ZA to derive directly an analytic form of the radial-velocity PDF, $P ( v_{\parallel}| r , t )$. We prefer the joint PDF, $P ( v_{\parallel} , v_{\perp} | r , t )$, which has a more simplified form and provides us with more straightforward information about, e.g., the origin of the negative skewness in the radial velocity (Section 5).

\subsection{Initial condition}

Here we determine the initial condition for $P ( v_{\parallel} , v_{\perp} | r , t )$. Although the same formulation was adopted in Seto \& Yokoyama (1998), we give a more complete description for our discussion in Sections 3 and 5. The basic assumption is that the initial density fluctuation is random Gaussian while the initial peculiar-velocity field is homogeneous, isotropic, random and vorticity-free.

Since the initial density fluctuation is random Gaussian, the corresponding PDF of the radial and transverse velocities is also Gaussian (Fisher 1995), 
\begin{equation}
P \left( v_{\parallel,i} , v_{\perp,i} | r_{i} , t_{i}\right) 
= \frac{1}{\sqrt{(2 \pi)^3 \sigma_{\parallel}^2(r_i) \sigma_{\perp}^4(r_i)}} 
\exp \left[ - \frac{1}{2} \left( \frac{v_{\parallel,i}^2}{\sigma_{\parallel}^2(r_i)} 
+ \frac{v_{\perp,i}^2}{\sigma_{\perp}^2(r_i)} \right) \right] .
\label{pdfini6}
\end{equation}
Thus we only have to obtain the dispersions of the initial pairwise velocities $\sigma_{\parallel}^2(r_i)$ and $\sigma_{\perp}^2(r_i)$. The averages $\langle v_{\parallel,i} \rangle$ and $\langle v_{\perp,i} \rangle$ do not have to be considered because they vanish in the limit $D_i \rightarrow 0$.

Since the initial peculiar-velocity field is homogeneous, the pairwise-velocity dispersions are derived from the one-point velocity dispersion and the two-point velocity correlation:
\begin{eqnarray}
\begin{array}{ll}
\sigma_{\parallel}^2(r_i) 
= 2 \langle u_{\parallel ,i}^2({\bmath x}_i) \rangle_s -
  2 \langle u_{\parallel ,i}({\bmath x}_i) u_{\parallel ,i}({\bf x}_i+{\bf r}_i) \rangle_s
   \\
\sigma_{\perp}^2(r_i) 
= 2 \langle u_{\perp ,i}^2({\bmath x}_i) \rangle_s -
  2 \langle u_{\perp ,i}({\bmath x}_i) u_{\perp ,i}({\bmath x}_i+{\bmath r}_i) \rangle_s.
\label{pdfini11}
\end{array}
\end{eqnarray}
Here $\langle \cdot \rangle_s$ denotes a spatial average and should not be confused with $\langle \cdot \rangle$ that denotes an average over galaxy pairs; $u_{\parallel ,i}$ and $u_{\perp ,i}$ are the radial and transverse components of the peculiar velocity ${\bmath u}_i$, respectively. We relate the velocity correlations to the power spectrum of the density fluctuation as follows (G\'orski 1988). Using a theory for a vector field that is homogeneous, isotropic, random and vorticity-free (Monin \& Yaglom 1975, section 12), we have
\begin{eqnarray}
\begin{array}{ll}
\left\langle 
u_{\parallel ,i}({\bmath x}_i) u_{\parallel ,i} ({\bmath x}_i + {\bmath r}_i) 
\right\rangle_s
= 2 \displaystyle 
\int^{\infty}_{0} \left[ j_{0} (k_ir_i) - 2 \displaystyle 
\frac{j_{1} (k_ir_i)}{k_ir_i} \right] {\cal E}_i(k_i) dk_i, \\
\left\langle 
u_{\perp ,i} ({\bmath x}_i) u_{\perp ,i} ({\bmath x}_i + {\bmath r}_i) 
\right\rangle_s
= 2 \displaystyle \int^{\infty}_{0}  
\frac{j_{1} (k_ir_i)}{k_ir_i} {\cal E}_i(k_i) dk_i. 
\label{pdfini1}
\end{array}  
\end{eqnarray}
Here $j_{0}$ and $j_{1}$ are first-kind spherical Bessel functions, 
\begin{equation}
j_{0}(x) = \displaystyle \frac{\sin x}{x}
\quad {\rm and} \quad
j_{1}(x) = \displaystyle \frac{\sin x}{x^2} - \frac{\cos x}{x},
\label{pdfini3}
\end{equation}
and ${\cal E}_i(k_i)$ is the power spectrum of the initial peculiar-velocity field, 
\begin{eqnarray}
\int {\cal E}_i(k_i) dk_i = \frac{1}{2} \left\langle \left| {\bmath u}_i({\bmath x}_i) \right|^2 
\right\rangle_s.
\label{pdfini5}
\end{eqnarray}
Using the linear perturbation, we relate the power spectrum of the initial peculiar-velocity field ${\cal E}_i(k_i)$ to that of the initial density fluctuation ${\cal P}_i(k_i)$. The linear peculiar-velocity field is described by the linear density field as 
\begin{eqnarray}
{\bmath u}_i({\bmath x}_i) = i \frac{\dot{D}_{i}}{D_{i}} \int \frac{{\bmath k}_i}{k_i^2}
\tilde{\delta}_i({\bmath k}_i) \exp(i {\bmath k}_i {\bmath x}_i) 
\frac{d {\bmath k}_i}{(2 \pi)^{3/2}},
\label{pdfini7}
\end{eqnarray}
where $\tilde{\delta}_i({\bmath k}_i)$ is the Fourier transform of the density contrast $\delta_i ({\bmath x}_i)$ (Peebles 1993, section 21). The power spectrum ${\cal P}_i(k_i) $ of the initial density fluctuation is  
\begin{eqnarray}
\left\langle \tilde{\delta}_i({\bmath k}_i) 
\tilde{\delta}_i({\bmath k}'_i)^{\ast} \right\rangle_s 
= {\cal P}_i (k_i) \left( 2 \pi \right)^{3} \delta ({\bmath k}_i - {\bmath k}'_i),
\label{pdfini9}
\end{eqnarray}
where $\ast$ denotes the complex conjugate. From equations (\ref{pdfini5})--(\ref{pdfini9}), the power spectrum ${\cal E}_i (k_i)$ of the initial peculiar-velocity field is obtained as 
\begin{eqnarray}
{\cal E}_i (k_i)
= 2 \pi \left( \frac{\dot{D}_{i}}{D_{i}} \right)^{2} 
{\cal P}_i (k_i).
\label{pdfini10}
\end{eqnarray}
Therefore, the radial- and transverse-velocity dispersions at the initial stage are 
\begin{eqnarray}
\begin{array}{ll}
\label{pdfini12}
\displaystyle \sigma_{\parallel}^2(r_i)
= \displaystyle \frac{8 \pi}{3} \left( \frac{\dot{D}_{i}}{D_{i}} \right)^{2} 
\int \left[ 1 - 3 j_{0} (k_ir_i) 
+ 6 \frac{j_{1} (k_ir_i)}{k_ir_i} \right] {\cal P}_{i} (k_i) dk_i
= \displaystyle \left( \frac{\dot{D}_{i}}{D_{i}} \right)^{2} R_{\parallel}^2 (r_i),
\\
\displaystyle \sigma_{\perp}^2 (r_i)
= \displaystyle \frac{8 \pi}{3} \left( \frac{\dot{D}_{i}}{D_{i}} \right)^{2} 
\int \left[ 1 - 3 \frac{j_{1} (k_ir_i)}{k_ir_i} \right] {\cal P}_{i} (k_i) dk_i
= \displaystyle \left( \frac{\dot{D}_{i}}{D_{i}} \right)^{2} R_{\perp}^2 (r_i).
\end{array}
\end{eqnarray}
The initial condition for $P ( v_{\parallel} , v_{\perp} | r , t)$ is determined by equations (\ref{pdfini6}) and (\ref{pdfini12}).

Since the peculiar velocity depends on the gravitational potential produced by the density fluctuation, equation (\ref{pdfini12}) includes the information that galaxies are about to cluster together. In fact, the radial velocity dispersion $\sigma_{\parallel}^2(r_i)$ reflects the two-point velocity correlations $\langle u_{\parallel ,i}({\bmath x}_i) u_{\parallel ,i}({\bf x}_i+{\bf r}_i) \rangle_s$. At a small separation $r_i$ where galaxies are about to move in the same direction, the correlation is positive. At a moderately large separation where galaxies are about to cluster from the opposite sides, the correlation is negative. At a very large separation, the correlation is absent.

\subsection{Power spectrum of initial density fluctuation}

The power spectrum of the initial density fluctuation ${\cal P}_i(k_i)$ is adopted from the standard cold dark matter model (Peebles 1993, section 25):
\begin{eqnarray}
{\cal P}_{i} (k_i) = \frac{B D_i^2 k_i}{\left\{ 1 + 
\left[ \alpha k_i + \left( \beta k_i \right)^{3/2} 
+ \left( \gamma k_i \right)^{2} \right]^{\nu} \right\}^{2/\nu}},
\label{inip}
\end{eqnarray}
where $\alpha = 25.6$ Mpc, $\beta = 12$ Mpc, $\gamma = 6.8$ Mpc and $\nu = 1.13$ for the Einstein-de Sitter universe with $H_0=50$ km s$^{-1}$ Mpc$^{-1}$ (Bond \& Efstathiou 1984; Efstathiou, Bond \& White 1992). The normalization factor $B$ is determined from the temperature fluctuation of the cosmic microwave background:
\begin{eqnarray}
B = \frac{12 \Omega_0^{-1.54} c^4}{5 \pi H_0^4} 
\left( \frac{{\cal Q}_{rms}}{T_0} \right)^2.
\end{eqnarray}
Here the quadrupole fluctuation amplitude ${\cal Q}_{rms}$ is set to be $9.5 \,{\rm \mu K}$ with the present-day temperature $T_0 = 2.73$ K. The same initial power spectrum was adopted in $N$-body simulations of Fisher et al. (1994) and Zurek et al. (1994), which are to be compared with our calculation, and also in the ZA calculation of Seto \& Yokoyama (1998). Other initial power spectra and cosmological models are studied in Appendix B.

Note that there are relations $v_{\parallel, i} \propto \dot{D}_i$, $v_{\perp, i} \propto \dot{D}_i$, $\sigma_{\parallel}(r_i) \propto \dot{D}_i$, $\sigma_{\perp}(r_i) \propto \dot{D}_i$ and ${\cal P}_i(k_i) \propto D_i^2$ in equations (\ref{pdf9}), (\ref{pdfini12}) and (\ref{inip}). Thus the initial condition for $P \left( v_{\parallel} , v_{\perp} | r , t\right)$ does not depend on the values of $D_i$ and $\dot{D}_i$.

The initial spectrum ${\cal P}_i(k_i)$ is used without any modification. Coles, Melott \& Shandarin (1993) suggested that ZA's performances are improved if small-scale fluctuations are removed from the initial spectrum by using a window function. Although this appears to be true in the calculation of the matter distribution, the amplitudes of pairwise velocity dispersions would be incorrectly underestimated as shown in Seto \& Yokoyama (1998). Therefore, we do not adopt this modification.

\section{APPLICABLE RANGE OF ZEL'DOVICH APPROXIMATION}

Galaxies in ZA continue to move straight, and pass away each other even if they once cluster together. Since actual clustering galaxies are bounded gravitationally, ZA is no longer valid after this so-called shell crossing. At a fixed time, ZA is not applicable to small separations where the shell crossing is statistically significant. In Section 3.1, we analytically estimate the separation range where ZA is applicable. In Section 3.2, we make a numerical estimation on the basis of time evolution of the two-point correlation function. The applicability in the velocity range is also discussed in Section 3.3.

\begin{figure}
  \begin{center}
  \includegraphics[height=8cm]{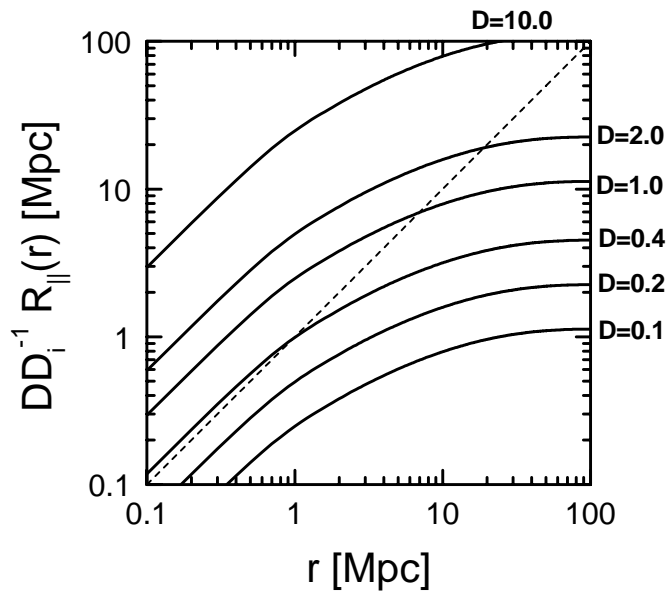}
  \caption{ Graphical representation of equation (\ref{zaapp4}) at $D = 0.1$, 0.2, 0.4, 1, 2 and 10 as a function of the separation $r$. The dashed line denotes the left-hand side of the equation. The solid lines denote the right-hand side. They are in units of Mpc. ZA is applicable over the separations where the dashed line is well above the solid line.}
  \end{center}
  \label{fig1}
\end{figure}

\subsection{Analytical estimation}

Let us consider a pair of galaxies that are approaching each other. If its initial separation is $r_{i}$, its typical initial radial velocity is estimated as $v_{\parallel ,i} \simeq -\sigma_{\parallel} (r_{i}) = -\dot{D}_{i} D_{i}^{-1} R_{\parallel} (r_{i})$. From equation (\ref{pdf1}), we have 
\begin{eqnarray}
r_{i} - r \simeq \displaystyle -\frac{D}{\dot{D_{i}}} v_{\parallel ,i} 
\simeq \displaystyle \frac{D}{D_{i}} R_{\parallel} (r_{i}).
\label{zaapp1}
\end{eqnarray}
The shell crossing occurs if $r = 0$. Thus ZA is valid as far as $r_{i} \gg D D_{i}^{-1} R_{\parallel} (r_{i})$. Since this condition implies $r \simeq r_{i}$, we replace $r_{i}$ with $r$ and obtain 
\begin{eqnarray}
r \gg \displaystyle \frac{D}{D_{i}}R_{\parallel} (r).
\label{zaapp4}
\end{eqnarray}
This is ZA's applicable range. Fig. 1 compares the right-hand side of equation (\ref{zaapp4}) with the left-hand side as a function of the separation $r$ at $D = 0.1$, 0.2, 0.4, 1, 2 and 10.  ZA is applicable over the separations where the dashed line (the left-hand side) is well above the solid line (the right-hand side). It is evident that the applicable range is progressively narrow with a progress of the time. At $D = 1$, ZA is applicable to $r \gg 10$ Mpc.

\begin{figure}
  \begin{center}
  \includegraphics[height=10cm]{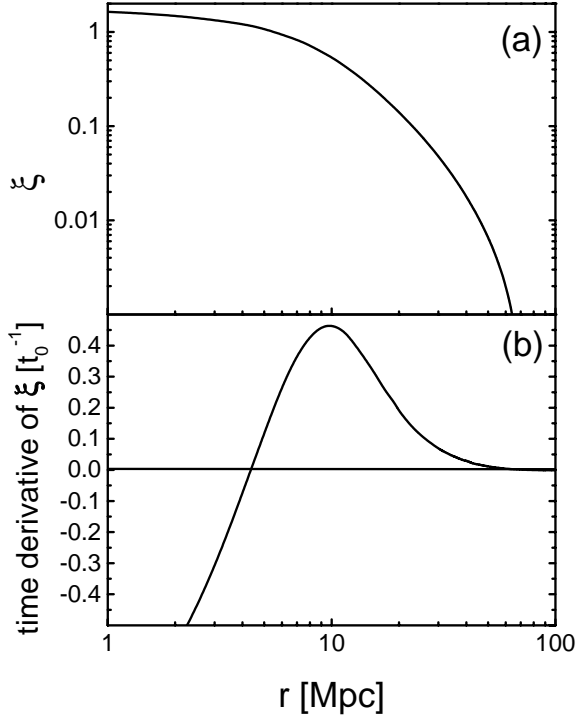}
  \caption{ Two-point correlation function $\xi$ ($a$) and its time derivative $\partial \xi / \partial t$ ($b$) at $D = 1$ in ZA. We show $\partial \xi / \partial t$ in units of $t_0^{-1}$, where $t_0$ is the present age of the universe. The abscissa is the separation $r$ in units of Mpc.}
  \end{center}
  \label{fig2}
\end{figure}

\subsection{Numerical estimation}

The shell crossing affects the time evolution of the two-point correlation function. This is because, at a given separation, galaxies become less clustered after the onset of the shell crossing. Using ZA, we numerically calculate the two-point correlation function $\xi$ and its time derivative $\partial \xi/\partial t$ at $D = 1$ (see equation \ref{norm}). The results are plotted as a function of the separation $r$ in Fig 2. The time growth $\partial \xi /\partial t$ of the two-point correlation function is maximal at $r \simeq 10 \,{\rm Mpc}$. This separation coincides with that for $r = D D_i^{-1} R_{\parallel} (r)$ (fig. 1 in Section 3.1). Thus ZA's applicable range is $r \gg 10$ Mpc at $D = 1$. Any discussion on ZA has to be restricted within this range.

\subsection{Applicability in velocity range}

Even if the separation $r$ is within ZA's applicable range, ZA is valid only for the velocity range
\begin{equation}
v_{\parallel} \ll \frac{\dot{D}}{D} r.
\end{equation}
For $v_{\parallel} > \dot{D} D^{-1} r > 0$, the initial radial velocity is negative as shown in equation (\ref{pdf9}). Such a galaxy pair has once clustered together with a negative radial velocity, experienced the shell crossing, and then turned to have a positive radial velocity. In the limit $|v_{\parallel}| \rightarrow \infty$, equations (\ref{pdfeq12})--(\ref{pdfini6}) and (\ref{pdfini12}) yield a Gaussian PDF,
\begin{equation}
P \left( v_{\parallel}, v_{\perp} | r , t \right)
\rightarrow
\frac{1}{\sqrt{(2 \pi R_{\infty} )^3}}
\left( \frac{D_i}{\dot{D}} \right) ^3
\exp \left[ - \frac{ v_{\parallel}^2 + v_{\perp}^2 }{2 R_{\infty}}
              \left( \frac{D_i}{\dot{D}} \right) ^2
     \right],
\end{equation}
where
\begin{equation}
R_{\infty} = \frac{8 \pi}{3} \int {\cal P}_{i} (k_i) dk_i.
\end{equation}
This is overestimation for the positive radial velocity because of the shell crossing as stated above.

\section{COMPARISON WITH $N$-BODY SIMULATIONS}

The radial-velocity PDFs are shown in Fig. 3 for $r = 1$, 5, 10 and 23 Mpc at $D = 0.1$, 0.2, 0.4, 1, 2 and 10 (solid lines). They are compared with the results of $N$-body simulations of Zurek et al. (1994) and Fisher et al. (1994) at $D = 1$ (dashed lines). For reference, we also show Gaussian PDFs (dotted lines).

\begin{figure}
  \begin{center}
  \includegraphics[height=12cm]{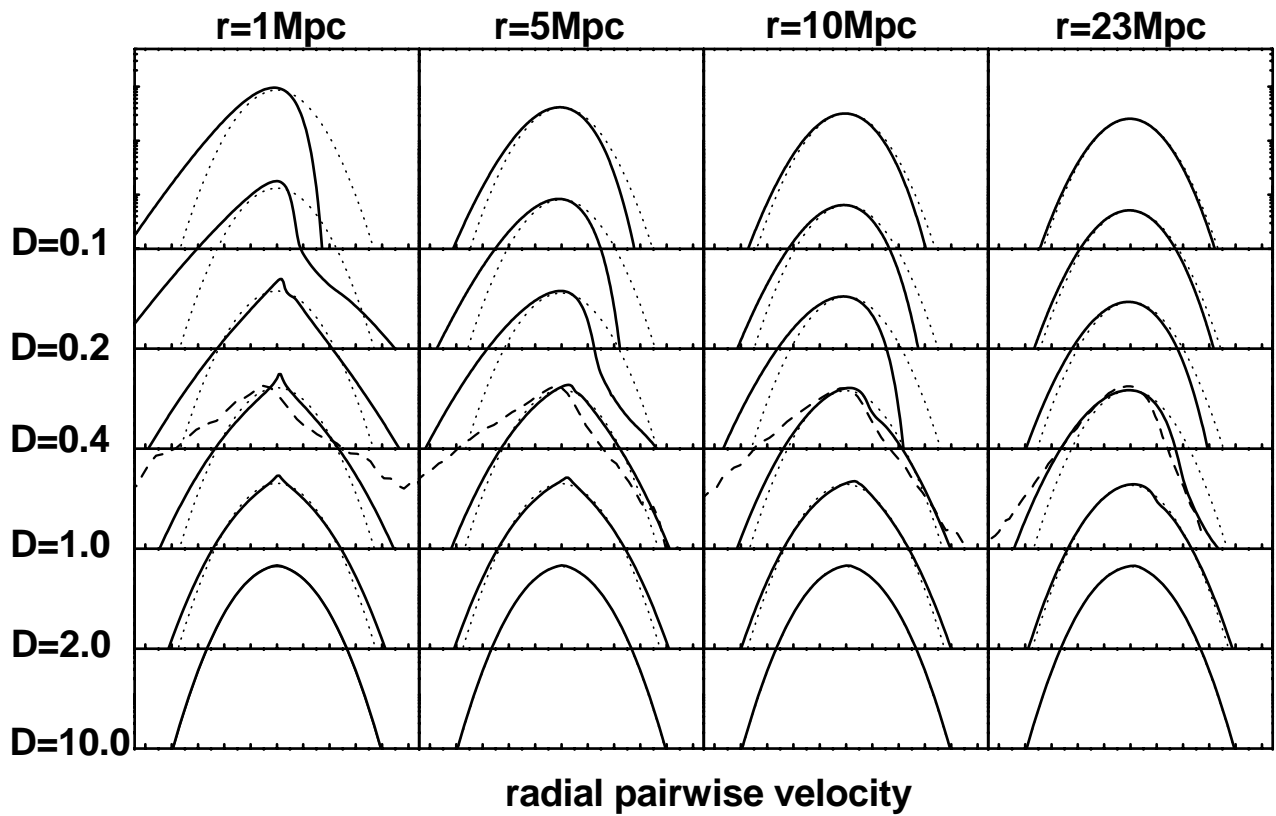}
  \caption{ Radial-velocity PDFs. {\it Solid lines}: ZA calculation at $D=0.1$, 0.2, 0.4, 1, 2 and 10 for $r = 1$, 5, 10 and 23 Mpc. {\it Dashed lines}: $N$-body simulations at $D = 1$ of Zurek et al. (1994) for $r =1$--2, 5--6 and 10--11 Mpc and of Fisher et al. (1994) for $r = 23$ Mpc. {\it Dotted lines}: Gaussian PDFs. The abscissa is in units of the radial-velocity dispersion $\sigma_{\parallel}(r)$ in ZA, which is 462 km s$^{-1}$ at $D = 1$ for $r = 23$ Mpc.}
  \end{center}
  \label{fig3}
\end{figure}

Fisher et al. (1994) presented the radial-velocity PDF at $D=1$ for $\xi = 0.1$, which corresponds to $r \simeq 23$ Mpc (fig. 2$a$). This separation is within ZA's applicable range. The PDF for $\xi = 0.1$ in the $N$-body simulation is in agreement with the PDF at $r = 23$ Mpc in ZA (fig. 3). Both the PDFs exhibit a negative average as well as negative skewness. They are surely due to gravitational clustering of galaxies.

The average $\langle v_{\parallel} \rangle$, standard deviation $\langle (v_{\parallel}- \langle v_{\parallel} \rangle )^2 \rangle^{1/2}$ and skewness $\langle (v_{\parallel}- \langle v_{\parallel} \rangle )^3 \rangle / \langle (v_{\parallel}- \langle v_{\parallel} \rangle )^2 \rangle ^{3/2}$ of the radial velocities are $-190$ km s$^{-1}$, 430 km s$^{-1}$ and $-0.69$, respectively, in the $N$-body simulation. They are $-133$ km s$^{-1}$, 462 km s$^{-1}$ and $-0.36$, respectively, in ZA. The velocity limit $\dot{D} D^{-1} r$ is 1150 km s$^{-1}$. The average and standard deviation in the $N$-body simulation are nearly the same as those in ZA. However, the skewness in the $N$-body simulation is somewhat different from that in ZA. The radial-velocity PDF in the $N$-body simulation has a more pronounced tail in the negative side. This is because gravitational acceleration of clustering galaxies is more significant than that assumed in ZA, which simply extrapolates the initial velocity ${\bmath v}_i$ (equation \ref{pdf1}).

For the separation $r = 23$ Mpc, the shell crossing becomes significant and ZA becomes invalid at $D \simeq 2$ (fig. 1). The PDF becomes symmetric (fig. 3). Eventually, almost all the galaxy pairs experience the shell crossing. The PDF becomes Gaussian as observed at $D = 10$.

Zurek et al. (1994) presented the radial-velocity PDF at $D=1$ for $r \simeq 1$, 5 and 10 Mpc. These separations are outside ZA's applicable range at $D = 1$ (figs 1 and 2). Fig. 3 shows the evolution of the PDF. At $D = 0.1$, ZA is valid. The PDF exhibits a negative average and negative skewness. With increasing $D$, the shell crossing becomes significant, ZA becomes invalid, and the PDF becomes symmetric. At $D = 1$, the PDF in ZA is quite different from that in the $N$-body simulation. While ZA yields $\langle v_{\parallel} \rangle = +8$, $+9$ and $-54$ km s$^{-1}$ for $r = 1$, 5 and 10 Mpc, respectively, the $N$-body simulation yields $\langle v_{\parallel} \rangle = -280$, $-355$ and $-276$ km s$^{-1}$.

Seto \& Yokoyama (1998) made a ZA calculation for $r = 1$--11 Mpc at $D = 1$. Since ZA and $N$-body simulations yield different radial-velocity PDFs, they concluded that gravitational clustering is not incorporated completely in ZA. Since the PDFs for $r \simeq 1$ and 5 Mpc in ZA and $N$-body simulations exhibit exponential tails, $\ln P ( v_{\parallel} | r , t ) \propto -|v_{\parallel}|$, they also concluded that simple kinematics as ZA results in these tails. However, the separations $r = 1$--11 Mpc at $D = 1$ are outside ZA's applicable range. The exponential tails at $r = 1$ and 5 Mpc in ZA are spurious and attributable to galaxies that have experienced the shell crossing for the positive side of the PDF and to galaxies that are still in gravitational infall for the negative side (see also Section 6).

\section{ORIGIN OF NEGATIVE SKEWNESS}

The negative average and negative skewness of the radial-velocity PDF originates in galaxy clustering, i.e., the infall $v_{\parallel} < 0$ dominates over the expansion $v_{\parallel} > 0$. We discuss how this process is incorporated in ZA. Since ZA is valid at $D D_i^{-1} R_{\parallel} (r) \ll r$, we separately study the velocity ranges $|v_{\parallel}| \ll \dot{D} D_i^{-1} R_{\parallel}$, $\dot{D} D_i^{-1} R_{\parallel} \ll |v_{\parallel}| \ll \dot{D} D^{-1} r $ and $\dot{D} D^{-1} r \ll |v_{\parallel}|$, where $\dot{D} D_i^{-1} R_{\parallel} = 495$ km s$^{-1}$ and $\dot{D} D^{-1} r = 1150$ km s$^{-1}$ at $D = 1$ for $r = 23$ Mpc.

First, we study the velocity range $\dot{D} D_i^{-1} R_{\parallel} \ll |v_{\parallel}| \ll \dot{D} D^{-1} r$. Since most of the galaxy pairs have relatively small transverse velocities $v_{\perp} \simeq \dot{D} \dot{D}_i^{-1} \sigma_{\perp} (r) \simeq \dot{D} D_i^{-1} R_{\perp} (r) \simeq \dot{D} D_i^{-1} R_{\parallel} (r)$, equations (\ref{pdf7}) and (\ref{pdf9}) are simplified as 
\begin{eqnarray}
 r_{i} &\simeq& r \left( 1 - \frac{D}{\dot{D}} \frac{v_{\parallel}}{r} \right),
\\
\label{eq26}
 v_{\parallel,i} &\simeq& \frac{\dot{D_{i}}}{\dot{D}} v_{\parallel}
\quad {\rm and} \quad
 v_{\perp,i} \simeq \frac{\dot{D_{i}}}{\dot{D}} v_{\perp}.
\end{eqnarray}
Thus the initial separation $r_i$ depends on the sign of the present radial velocity $v_{\parallel}$:
\begin{equation}
 r_{i} (v_{\parallel} = +V) < r_{i} (v_{\parallel} = - V),
\label{pdfana9} 
\end{equation}
where $V > 0$. For the initial separation $r_i$, the radial-velocity dispersion $\sigma_{\parallel}(r_i)$ has the following dependence as shown in Fig. 4$a$:
\begin{equation}
\label{isigma}
\sigma_{\parallel}(r_i') < \sigma_{\parallel}(r_i'') 
\quad {\rm for} \quad r_i' < r_i'' \la 100\ {\rm Mpc}.
\end{equation}
The initial PDF, $P ( v_{\parallel,i} , v_{\perp,i} | r_i , t_i )$, is Gaussian with the standard deviation $\sigma_{\parallel}(r_i)$. Equation (\ref{eq26}) implies $|v_{\parallel,i}| \gg \sigma_{\parallel}(r_i)$. For such an argument that exceeds the standard deviation, a Gaussian function with a larger standard deviation yields a larger value as shown in Fig. 4$b$. Hence we expect 
\begin{equation}
P \left( v_{\parallel,i} , v_{\perp,i} | r_i' , t_i \right) < P \left( v_{\parallel,i} , v_{\perp,i} | r_i'' , t_i \right) \quad {\rm for} \quad r_i' < r_i'',
\end{equation}
which leads to 
\begin{equation}
\label{pdfana12}
P \left( v_{\parallel}=+V , v_{\perp} | r , t \right) < P \left( v_{\parallel}=-V , v_{\perp} | r , t \right).
\end{equation}
Thus the infall $v_{\parallel} < 0$ dominates over the expansion $v_{\parallel} > 0$ in this velocity range. The PDF is accordingly asymmetric.

\begin{figure}
  \begin{center}
  \includegraphics[height=10cm]{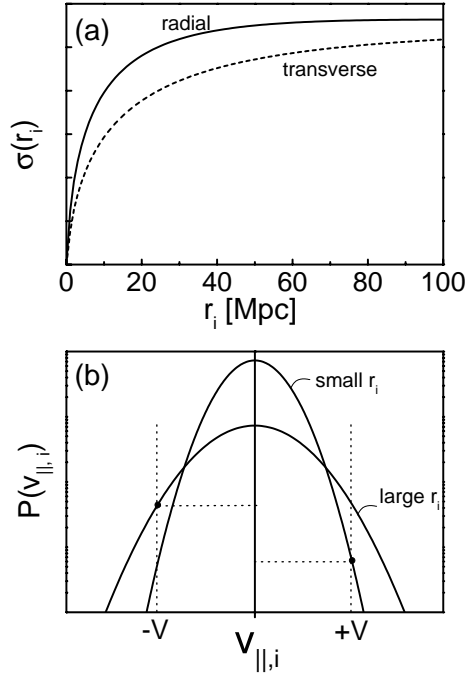}
  \caption{ ($a$) Dependence of the initial pairwise-velocity dispersions $\sigma_{\parallel}(r_i)$ ({\it solid line}) and $\sigma_{\perp}(r_i)$ ({\it dashed line}) on the initial separation $r_i$. The dispersions are in arbitrary units. The separation is in units of Mpc. ($b$) Two Gaussian PDFs with different standard deviations.}
  \end{center}
  \label{fig4}
\end{figure}

Second, we study the velocity range $|v_{\parallel}| \ll \dot{D} D_{i}^{-1} R_{\parallel}$. Since the absolute value of $v_{\parallel}$ tends to be much smaller than $v_{\perp}$, the sign of $v_{\parallel}$ is not important to the values of $r_i$ and $v_{\parallel ,i}$ in equations (\ref{pdf7}) and (\ref{pdf9}). We expect $P ( v_{\parallel}=+V , v_{\perp} | r , t ) \simeq P ( v_{\parallel}=-V , v_{\perp} | r , t )$. The radial velocities among galaxies with small peculiar velocities retain the initial Gaussian character. These galaxies have not moved significantly from their initial positions.

Third, we study the velocity range $|v_{\parallel}| \gg \dot{D} D^{-1} r$. Equations (\ref{pdf7}) and (\ref{pdf9}) yield $r_i \simeq D \dot{D}^{-1} |v_{\parallel}|$ and hence $v_{\parallel ,i} \simeq -\dot{D_i} \dot{D}^{-1} |v_{\parallel}|$, which in turn yields $P ( v_{\parallel}=+V, v_{\perp} | r , t ) \simeq P (v_{\parallel}=-V, v_{\perp} | r , t )$. This is due to shell crossing as discussed in Section 3.3. For the real universe, the radial-velocity PDF is expected to be asymmetric in this velocity range.

Therefore, the negative average and negative skewness in ZA stem from the inequality $P ( v_{\parallel}=+V , v_{\perp} | r , t ) < P ( v_{\parallel}=-V , v_{\perp} | r , t )$ in the velocity range $\dot{D} D_i^{-1} R_{\parallel} \ll |v_{\parallel}| \ll \dot{D} D^{-1} r$. The existence of this velocity range is assured by the condition for ZA's applicability, $D D_i^{-1} R_{\parallel} \ll r$. Whenever ZA is applicable, its radial-velocity PDF has a negative average and negative skewness.

The dependence of the radial-velocity dispersion $\sigma_{\parallel}(r_i)$ on the separation $r_i$ in equation (\ref{isigma}) is also an assured behavior. As discussed in Section 2.2, the radial-velocity dispersion $\sigma_{\parallel}(r_i)$ reflects the difference between the one-point velocity dispersion $\langle u_{\parallel ,i}^2({\bmath x}_i) \rangle_s$ and the two-point velocity correlation $\langle u_{\parallel ,i}({\bmath x}_i) u_{\parallel ,i}({\bf x}_i+{\bf r}_i) \rangle_s$. They are identical at $r_i = 0$. With an increase of the separation $r_i$, the two-point correlation decreases slowly in accordance with motions of galaxies that are about to move coherently and cluster together.

\section{DISCUSSION ON RELEVANT MODELS}

The pairwise velocity was studied by Juszkiewicz et al. (1998) using a second-order Eulerian perturbation. At the separation of 20.8 Mpc, the average, standard deviation and skewness of the radial velocities are $-200$ km s$^{-1}$, 430 km s$^{-1}$ and $-1.00$, respectively. They are $-190$ km s$^{-1}$, 440 km s$^{-1}$ and $-0.75$, respectively, in the $N$-body simulation under the same conditions. The agreement is better than it is in ZA (Section 4), probably because ZA does not conserve the momentum (Juszkiewicz et al. 1998). This could be a disadvantage of ZA, in addition to the problem of shell crossing. The second-order Eulerian perturbation has its own disadvantage that the formulation is too complicated to yield the radial-velocity PDF. Thus ZA and the Eulerian perturbation are complementary. Juszkiewicz et al. (1998) showed that $\langle (v_{\parallel}- \langle v_{\parallel} \rangle )^3 \rangle$ approximately scales as $\langle v_{\parallel} \rangle \langle (v_{\parallel}- \langle v_{\parallel} \rangle )^2 \rangle$. The negative skewness is induced by the negative average that reflects gravitational clustering of galaxies as discussed in Section 5.

The Lagrangian and Eulerian perturbations are useless in the strongly nonlinear regime, $\xi(r) \gg 1$ and accordingly $r \ll 10$ Mpc at $D = 1$, for which $N$-body simulations have shown that the radial-velocity PDF has an exponential tail (Section 4). We favor the model of Sheth (1996) and Diaferio \& Geller (1996). The exponential tails were reproduced by averaging relative velocities of galaxy pairs belonging to the same galaxy clusters. These clusters were assumed to be virialized as well as isothermal and have Gaussian velocity distributions with various dispersions according to the theory of Press \& Schechter (1974). The PDF is symmetric, i.e., $\langle v_{\parallel} \rangle = 0$. This is because there was assumed to exist no gravitational clustering. Since the $N$-body simulation yields $\langle v_{\parallel} \rangle < 0$ for $r = 1$ and 5 Mpc (Section 4), we consider that gravitational clustering still exists in practice at these separations.

\section{CONCLUSION}

The radial pairwise velocities of galaxies exhibit a negative average as well as negatively skewed pronounced tails in the quasi-nonlinear regime. To understand their origin, we have used ZA, an approximation for gravitational clustering, and analytically studied the radial-velocity PDF.

We have estimated for the first time the separation range where ZA is applicable. For the cold dark matter model of the Einstein-de Sitter universe with $H_0=50$ km s$^{-1}$ Mpc$^{-1}$, the applicable range is $r \gg 10$ Mpc at $D=1$ (Section 3). Any discussion on ZA has to be restricted within this range.

We have compared our ZA calculation with the result of the $N$-body simulation done under the same initial condition by Fisher et al. (1994). ZA reproduces successfully the negative average and negative skewness of the radial-velocity PDF (Section 4).

The negative average and negative skewness originate in galaxy clustering according to the initial gravitational potential, i.e., the infall $v_{\parallel} < 0$ dominates over the expansion $v_{\parallel} > 0$, as pointed out by Juszkiewicz et al. (1998). This information is contained in the dependence of the initial velocity dispersion on the initial separation (Section 5).

The radial-velocity PDF in $N$-body simulations has a pronounced tail in the negative side. Since ZA cannot fully reproduce this property, we have attributed it to gravitational acceleration of clustering galaxies that is not fully incorporated in ZA (Section 4). To confirm this conclusion, we would require higher-order approximations, e.g., post-ZA and post-post-ZA. They use not only the initial velocity ${\bmath v}_i$ but also the initial acceleration $\dot{{\bmath v}}_i$ and so on (Bouchet et al. 1995; see also Yoshisato et al. 1998). The approximations Pad\'e-post-ZA and Pad\'e-post-post-ZA developed by Matsubara, Yoshisato \& Morikawa (1998) would be also useful.

\section*{Acknowledgments}

This paper is in part the result of AY's research toward fulfillment of the requirements of the Ph. D. degree at Ochanomizu University. The authors are grateful to the referee for helpful comments.

\appendix

\section{TIME EVOLUTION OF PROBABILITY DISTRIBUTION FUNCTION}

Let us expand the joint PDF, $P ( v_{\parallel}' , v_{\perp}' | r' , t + dt ) $, around the time $t$ using ZA (equations \ref{pdf1} and \ref{pdfeq10}):
\begin{eqnarray}
\begin{array}{lll}
P \left( v_{\parallel}' , v_{\perp}' | r' , t + dt \right) 
\delta v_{\parallel}' 2 \pi v_{\perp}' \delta v_{\perp}' 4 \pi r'^2 \delta r'
\\ \quad
= \left[ P \left( v_{\parallel} , v_{\perp} | r , t \right) + 
 \displaystyle \frac{\partial P}{\partial v_{\parallel}} 
\left( \frac{\ddot{D}}{\dot{D}} v_{\parallel} 
  + \frac{v_{\perp}^2}{r} \right) dt 
 + \displaystyle \frac{\partial P}{\partial v_{\perp}}
\left( \frac{\ddot{D}}{\dot{D}} v_{\perp} 
  - \frac{v_{\perp} v_{\parallel}}{r} \right) dt \right.
\\ \left. \qquad \qquad
 + \displaystyle \frac{\partial P}{\partial r} v_{\parallel} dt 
 + \displaystyle \frac{\partial P}{\partial t} dt \right] 
\left( 1 + 3 \displaystyle \frac{\ddot{D}}{\dot{D}} dt \right) 
\delta v_{\parallel} 2 \pi v_{\perp} \delta v_{\perp} 4 \pi r^2 \delta r.
\end{array}
\label{pdfyeq1}
\end{eqnarray}
From comparison between equations (\ref{pdfeq1}) and (\ref{pdfyeq1}), we obtain the time-evolution equation of the joint PDF as 
\begin{eqnarray}
\displaystyle \frac{\partial P}{\partial t} 
= - \displaystyle \frac{\ddot{D}}{\dot{D}} 
\left( v_{\parallel} \displaystyle 
\frac{\partial P}{\partial v_{\parallel}} 
 + v_{\perp} \displaystyle \frac{\partial P}{\partial v_{\perp}} 
+3 P \right)
- \displaystyle \frac{v_{\perp}}{r} 
\left( v_{\perp} \displaystyle 
\frac{\partial P}{\partial v_{\parallel}} 
 - v_{\parallel} \displaystyle \frac{\partial P}{\partial v_{\perp}} \right)
 - v_{\parallel} \displaystyle \frac{\partial P}{\partial r}.
\label{pdfyeq2}
\end{eqnarray}
This equation is equivalent to a collisionless Boltzmann equation because ZA ignores encounters among the individual galaxies. The first, second and third terms represent the acceleration, rotation and translation of the galaxies, respectively. However, $(v_{\parallel}, v_{\perp} , r)$ do not constitute the generalized coordinates. This fact yields the third term $3P$ in the parenthesis of the first term (see equation \ref{pdfeq10}).

The integration of equation (\ref{pdfyeq2}) by $d v_{\parallel} 2 \pi v_{\perp} d v_{\perp}$ yields the evolution of the two-point correlation function $\xi (r,t)$,
\begin{eqnarray}
\displaystyle \frac{\partial}{\partial t} \left( 1 + \xi (r,t) \right) = 
\displaystyle \frac{\partial \xi (r,t)}{\partial t} 
= - \displaystyle \frac{1}{r^2} \frac{\partial}{\partial r} 
 \left[ r^2 \left( 1 + \xi (r,t) \right) 
 \left\langle v_{\parallel} \right\rangle \right],
\label{pdfyeq5}
\end{eqnarray}
which corresponds to a continuity equation. This is a reasonable result because the galaxy number is conserved as in equation (\ref{pdfeq1}).

\section{PAIRWISE VELOCITIES IN OTHER COSMOLOGICAL MODELS}

Thus far we have assumed the Einstein-de Sitter universe with a Hubble constant $H_0=50$ km s$^{-1}$ Mpc$^{-1}$. The application to other cosmological models is straightforward. We only have to know the values of $D$ and $\dot{D}$ and substitute them to equations (\ref{pdfeq12})--(\ref{pdf9}). The functional forms of $D$ are summarized in Suto (1993).

\begin{figure}
  \begin{center}
  \includegraphics[height=9cm]{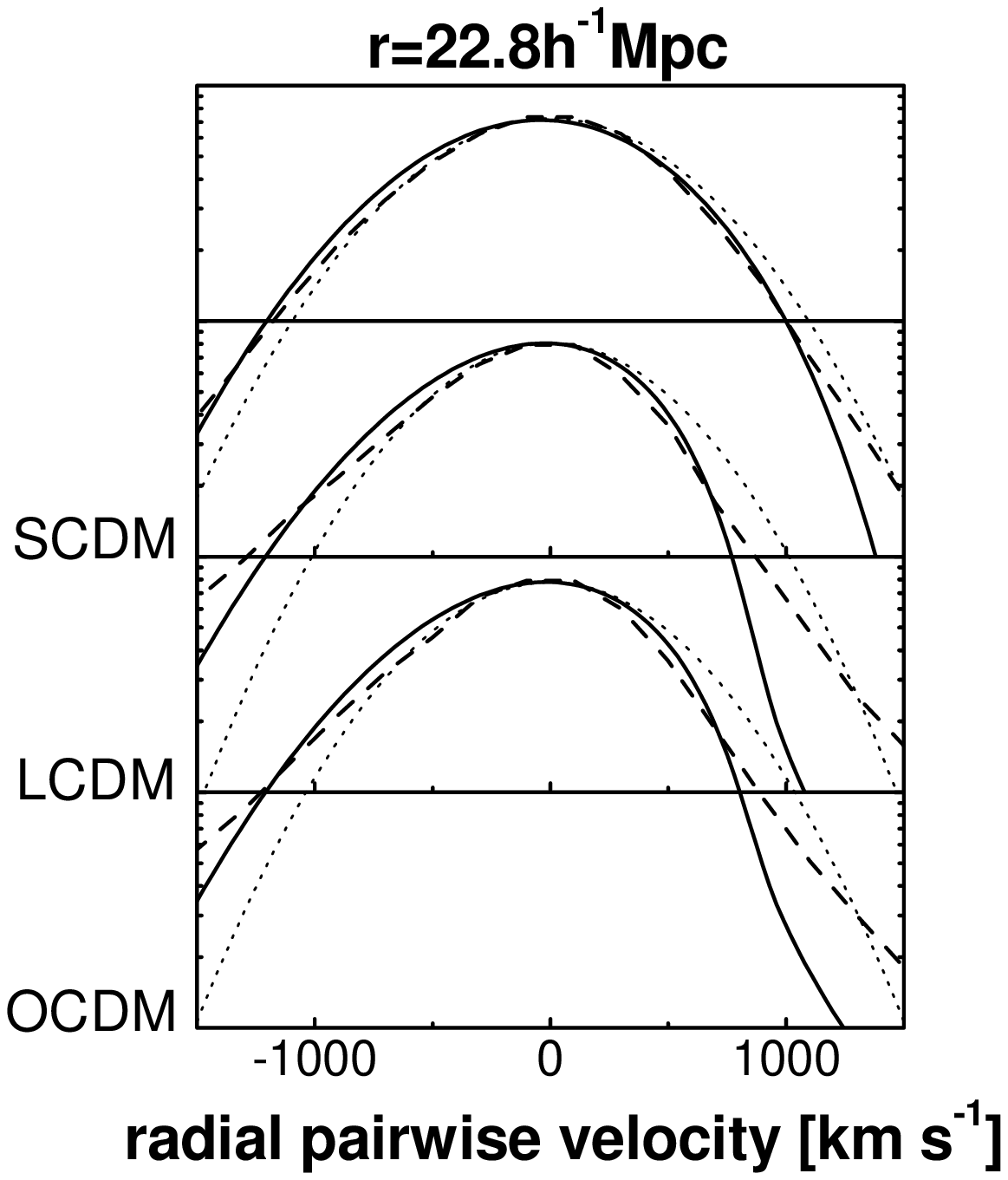}
  \caption{Radial-velocity PDFs for SCDM, LCDM and OCDM at $r = 22.8$ $h^{-1}$ Mpc. {\it Solid lines}: ZA calculation. {\it Dashed lines}: $N$-body simulation of Magira et al. (2000). {\it Dotted lines}: Gaussian PDFs with the same dispersions $\sigma_{\parallel}(r)$ as ZA's PDFs. The abscissa is in units of km s$^{-1}$.}
  \end{center}
  \label{figC1}
\end{figure}

We compare ZA with $N$-body simulations of Magira et al. (2000, see also Suto et al. 1999). They considered three models, SCDM ($\Omega_0 = 1.0$, $\lambda_0 = 0.0$, $h = 0.50$ and $\Gamma = 0.50 $), LCDM ($\Omega_0 = 0.3$, $\lambda_0 = 0.7$, $h = 0.70$ and $\Gamma = 0.21$) and OCDM ($\Omega_0 = 0.3$, $\lambda_0 = 0.0$, $h = 0.83$ and $\Gamma = 0.25$). Here $\Omega_0$ is the density parameter, $\lambda_0$ is the normalized cosmological constant, and $h$ is defined as $H_0$/(100 km s$^{-1}$ Mpc$^{-1}$). Thus SCDM is the Einstein-de Sitter universe, while LCDM is a flat universe with a cosmological constant and OCDM is an open universe. The power spectrum of the initial density fluctuation  ${\cal P}_i(k_i)$ was from the cold dark matter model of Bardeen et al. (1986, see also Sugiyama 1995):
\begin{equation}
{\cal P}_i(k_i) = B D_i^2 k_i 
\frac{\ln^2(1+2.34q)}{(2.34q)^2}
\left[ 1+3.89q+(16.1q)^2+(5.46q)^3+(6.71q)^4 \right]^{-1/2},
\end{equation}
where $q = k_i/\Gamma h$. The normalization factor $B$ was determined by comparing the radial-velocity dispersion at infinity $\sigma_{\parallel}(r \rightarrow \infty)$ in the linear theory of equation (\ref{pdfini12}) with that in the $N$-body simulation (591, 606 and 603 km s$^{-1}$ for SCDM, LCDM and OCDM). Fig. B1 shows the radial-velocity PDF at 22.8 $h^{-1}$ Mpc. This separation is within ZA's applicable range. ZA and the $N$-body simulation are in satisfactory agreement, except for the tails of the PDF that are pronounced in the $N$-body simulation. In SCDM, the average, standard deviation and skewness are $-66$ km s$^{-1}$, 589 km s$^{-1}$ and $-0.22$ in the $N$-body simulation, while they are $-88$ km s$^{-1}$, 551 km s$^{-1}$ and $-0.18$ in ZA. In LCDM, the average, standard deviation and skewness are $-153$ km s$^{-1}$, 616 km s$^{-1}$ and $-0.51$ in the $N$-body simulation, while they are $-155$ km s$^{-1}$, 496 km s$^{-1}$ and $-0.38$ in ZA. In OCDM, the average, standard deviation and skewness are $-127$ km s$^{-1}$, 618 km s$^{-1}$ and $-0.40$ in the $N$-body simulation, while they are $-136$ km s$^{-1}$, 512 km s$^{-1}$ and $-0.28$ in ZA. 


\label{lastpage}

\end{document}